\documentstyle[aps,epsf,rotate,multicol]{revtex}

\begin{document}

\draft

\title{Classes of behavior of small-world networks}

\author{Lu\'{\i}s A. Nunes Amaral, Antonio Scala, Marc Barth\'el\'emy,
and H. Eugene Stanley}

\address{Center for Polymer Studies and Department of Physics \\ 
        Boston University, MA 02215, USA}


\maketitle

\vspace*{0.5cm}

\begin{abstract}

Small-world networks are the focus of recent interest because
they appear to circumvent many of the limitations of either random
networks or regular lattices as frameworks for the study of
interaction networks of complex systems. Here, we report an empirical
study of the statistical properties of a variety of diverse real-world
networks.  We present evidence of the occurrence of three classes of
small-world networks: (a) {\it scale-free\/} networks, characterized
by a vertex connectivity distribution that decays as a power law; (b)
{\it broad-scale\/} networks, characterized by a connectivity
distribution that has a power-law regime followed by a sharp cut-off;
(c) {\it single-scale\/} networks, characterized by a connectivity
distribution with a fast decaying tail.  Moreover, we note for the
classes of {\it broad-scale\/} and {\it single-scale\/} networks that
there are constraints limiting the addition of new links.  Our results
suggest that the nature of such constraints may be the controlling
factor for the emergence of different classes of networks. 

\end{abstract}

\begin{multicols}{2}


Disordered networks, such as small-world networks are the focus of
recent interest because of their potential as models for the
interaction networks of complex systems
\cite{Watts98,Watts99,Barthelemy99,Barabasi99a,Barabasi99b,Fernandez}.
Specifically, neither random networks nor regular lattices appear to
be an adequate framework within which to study ``real-world'' complex
systems \cite{Kochen89} such as chemical-reaction networks
\cite{Alon99}, neuronal networks~\cite{Watts98}, food-webs
\cite{Pimm91,Paine92,McCann98}, social networks
\cite{Wasserman94,Axtell99a,Axtell99b}, scientific-collaboration
networks \cite{vanRaan90}, and computer networks
\cite{Barabasi99a,Huberman98,Huberman99}.

Small-world networks~\cite{Watts98,Watts99} ---which emerge as the
result of randomly replacing a fraction $p$ of the links of a
$d$-dimensional lattice with new random links--- interpolate between
the two limiting cases of a regular lattice ($p=0$) and a random graph
($p=1$).  A ``small-world'' network is characterized by the properties
(i) the local neighborhood is preserved ---as for regular
lattices~\cite{Watts98,Watts99}---, and (ii) the diameter of the
network, quantified by average shortest distance between two
vertices~\cite{Leeuwen90}, increases logarithmically with the number
of vertices $n$ ---as for random graphs~\cite{Bollobas85}.  The latter
property gives the name ``small-world'' to these networks, as it is
possible to connect any two vertices in the network through just a few
links while the local connectivity would suggest the network to be of
finite dimensionality.

The structure of small-world networks and of real networks has been
probed through the calculation of their diameter as a function of
network size~\cite{Watts98}.  In particular, networks such as (a) the
electric-power grid for Southern California, (b) the network of
movie-actor collaborations, and (c) the neuronal network of the worm
{\it C.~Elegans\/}, appear to be small-world networks~\cite{Watts98}.
Further, it was proposed~\cite{Barabasi99b} that these three networks,
the world-wide web \cite{Barabasi99a}, and the network of citations of
scientific papers \cite{Seglen92,Redner98} are scale-free ---that is,
they have a distribution of connectivities that decays with a
power-law tail.

Scale-free networks emerge in the context of a growing network in
which new vertices connect preferentially to the more highly connected
vertices in the network~\cite{Barabasi99b}.  Scale free networks are
still small-world networks because (i) they have clustering
coefficients much larger than random networks \cite{Watts98,Watts99},
and (ii) their diameter increases logarithmically with the number of
vertices $n$~\cite{Barabasi99b}.


Here, we address the question of the conditions under which disordered
networks are scale-free through the analysis of several networks in
social, economic, technologic, biologic, and physical systems.  We
identify a number of systems for which there is a single scale for the
connectivity of the vertices.  For all these networks there are
constraints limiting the addition of new links.  Our results suggest
that such constraints may be the controlling factor for the emergence
of scale-free networks.

\begin{figure}
\centerline{\bf FIGURE 1}
\vspace*{1cm}
\centerline{
\epsfysize=0.4\columnwidth{{\epsfbox{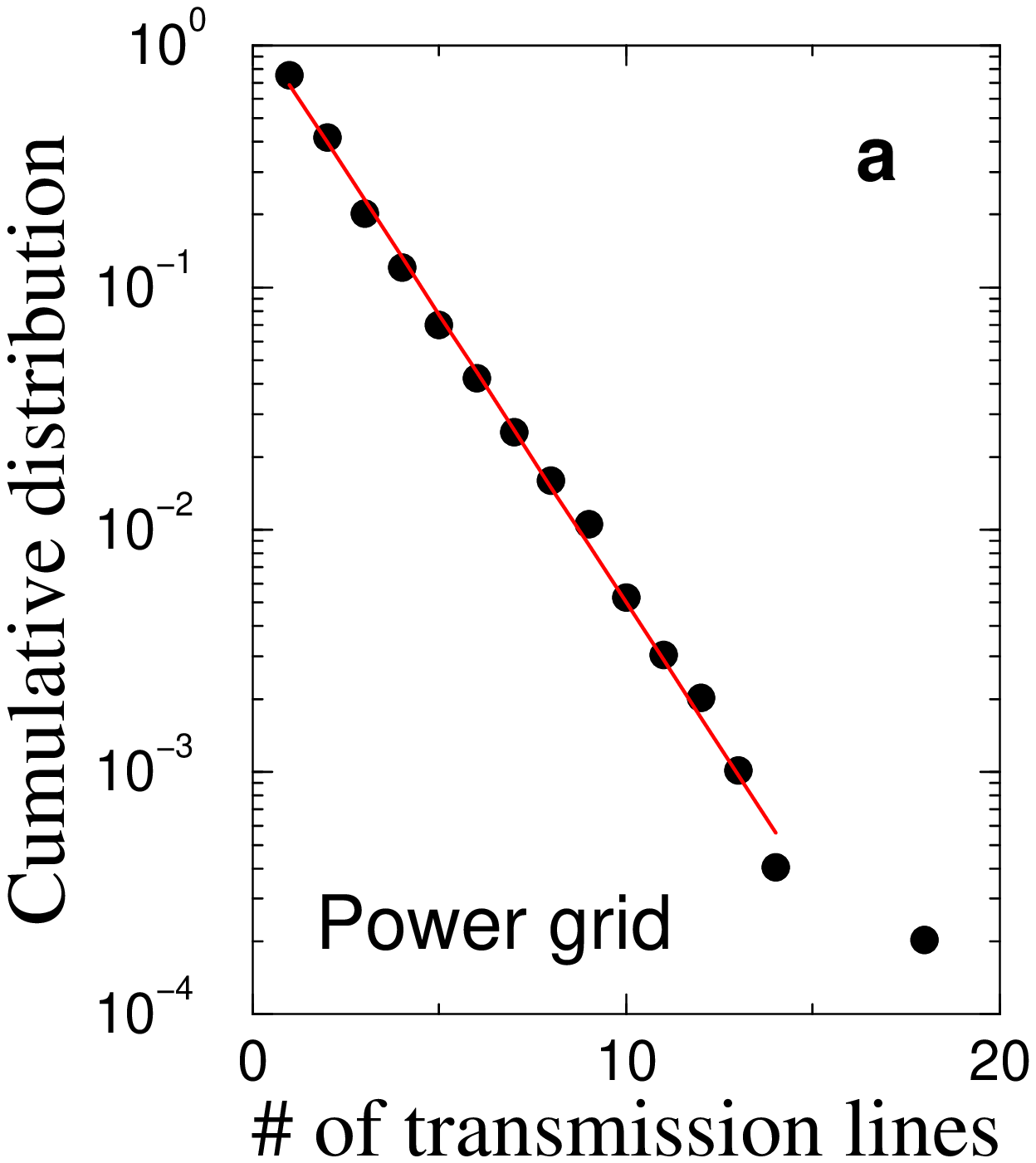}}}
\hspace*{-1.5cm}
\epsfysize=0.4\columnwidth{{\epsfbox{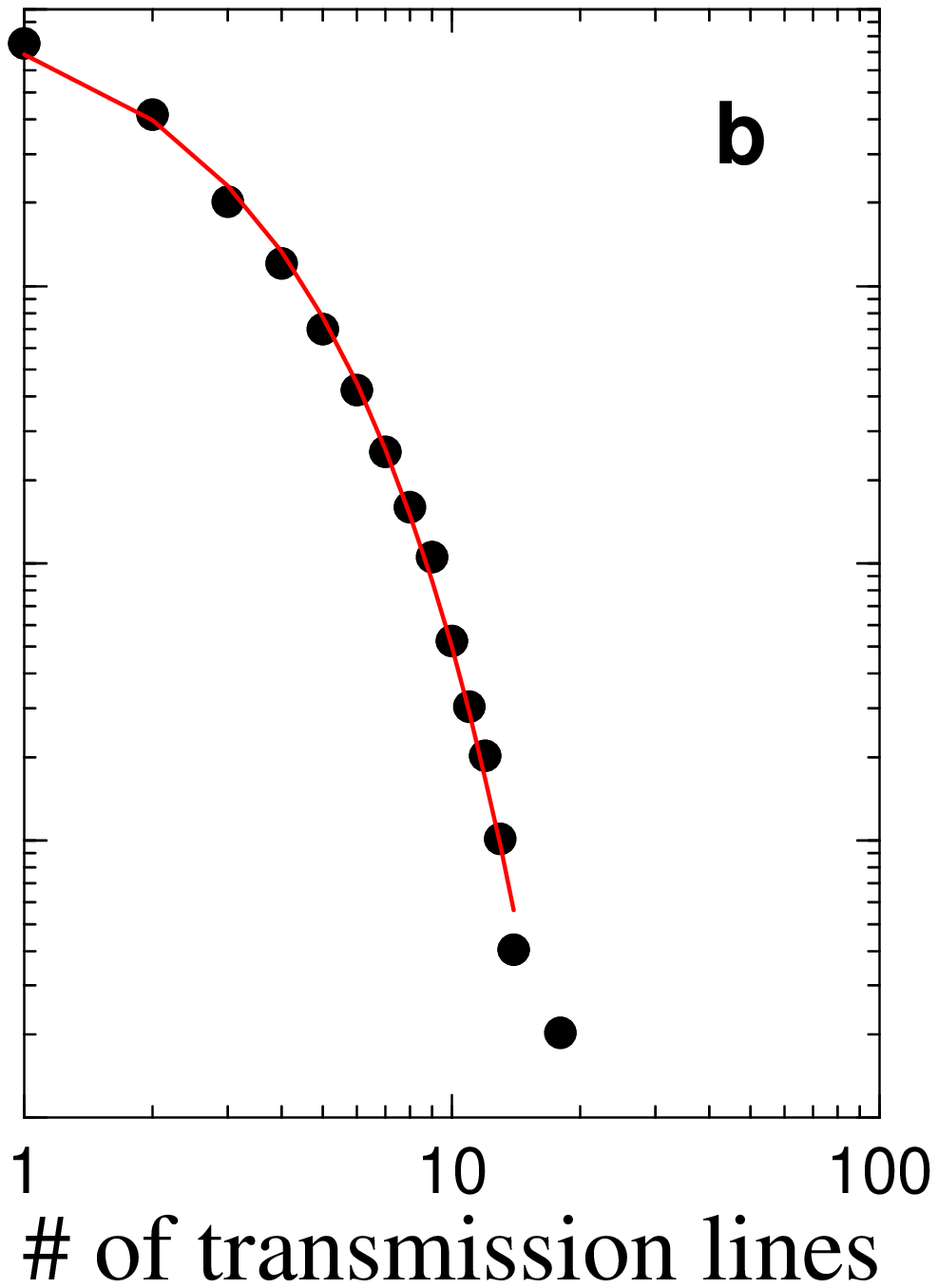}}}
}
\centerline{
\epsfysize=0.4\columnwidth{{\epsfbox{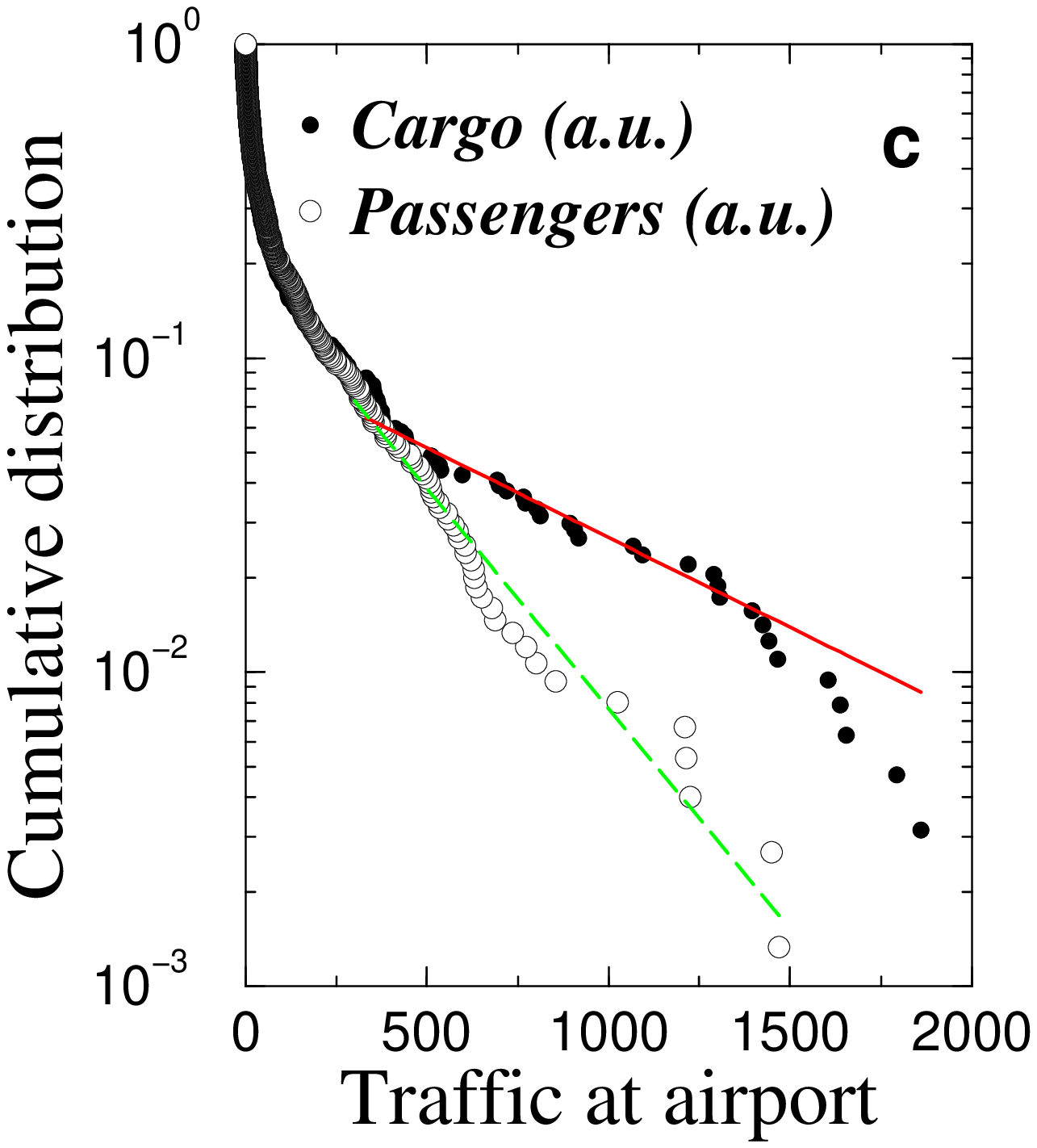}}}
\hspace*{-1.5cm}
\epsfysize=0.4\columnwidth{{\epsfbox{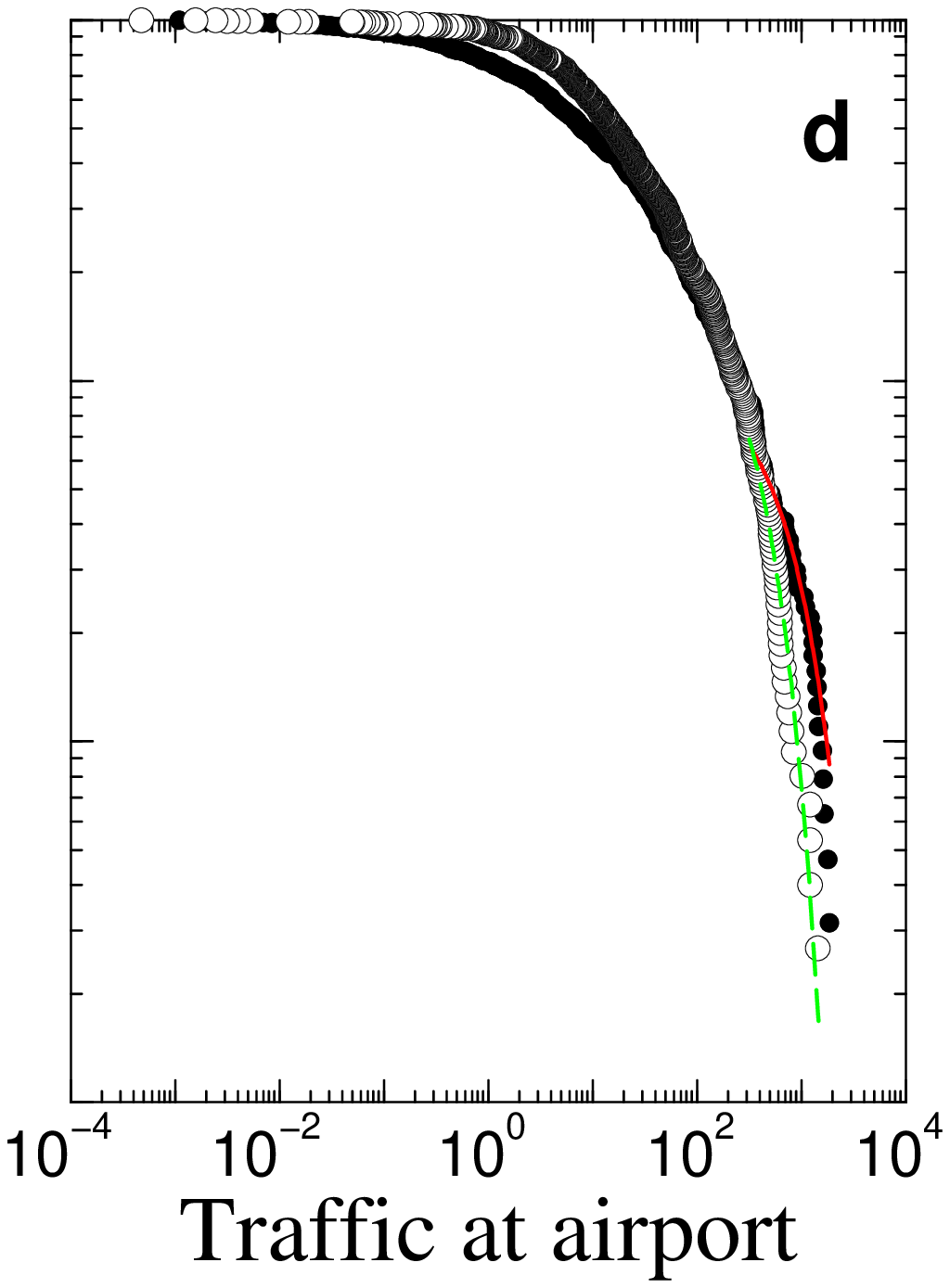}}}
}
\vspace*{.5cm}
\caption{ Technologic and economic networks. {\bf a} Linear-log plot
of the cumulative distribution of connectivities for the
electric-power grid of Southern California~\protect\cite{Watts98}.
For this type of plot, the distribution falls on a straight line
indicating an exponential decay of the distribution of connectivities.
The full line, which is an exponential fit to the data, displays good
agreement with the data. {\bf b} Log-log plot of the cumulative
distribution of connectivities for the electric-power grid of Southern
California.  If the distribution would have a power law tail then it
would fall on a straight line in a log-log plot.  Clearly, the data
reject the hypothesis of power-law distribution for the
connectivity. {\bf c} Linear-log plot of the cumulative distribution
of traffic at the world's largest airports for two measures of
traffic, cargo and number of passengers.  The network of world
airports is a small-world network: one can connect any two airports in
the network by only 1-5 links.  To study the distribution of
connectivities of this network, we assume that, for a given airport,
cargo and number of passengers are proportional to the number of
connections of that airport with other airports.  The data are
consistent with a decay of the distribution of connectivities for the
network of world airports that decays exponentially or faster.  The
full line is an exponential fit to the cargo data for values of
traffic between 500 and 1500. For values of traffic larger than 1500,
the distribution appears to decay even faster than an exponential. The
long-dashed line is an exponential to the passenger data for values of
traffic between 500 and 1500.  {\bf d} Log-log plot of the cumulative
distribution of traffic at the world's largest airports.  This plot
confirms that the tails of the distributions decay faster than a
power law would. }
\label{f.economic}
\end{figure}

First, we consider two examples of technologic and economic networks:
(i) the electric-power grid of Southern California~\cite{Watts98}, the
vertices being generators, transformers and substations and the links
high-voltage transmission lines, and (ii) the network of world
airports~\cite{ACI}, the vertices being the airports and the links
non-stop connections.  Figure~\ref{f.economic} shows the connectivity
distribution for these two examples.  It is visually apparent that
neither case has a power-law regime, and that both have exponentially
decaying tails, implying that there is a single scale for the
connectivity $k$.

\begin{figure}
\centerline{\bf FIGURE 2}
\vspace*{1cm}
\centerline{
\epsfysize=0.4\columnwidth{{\epsfbox{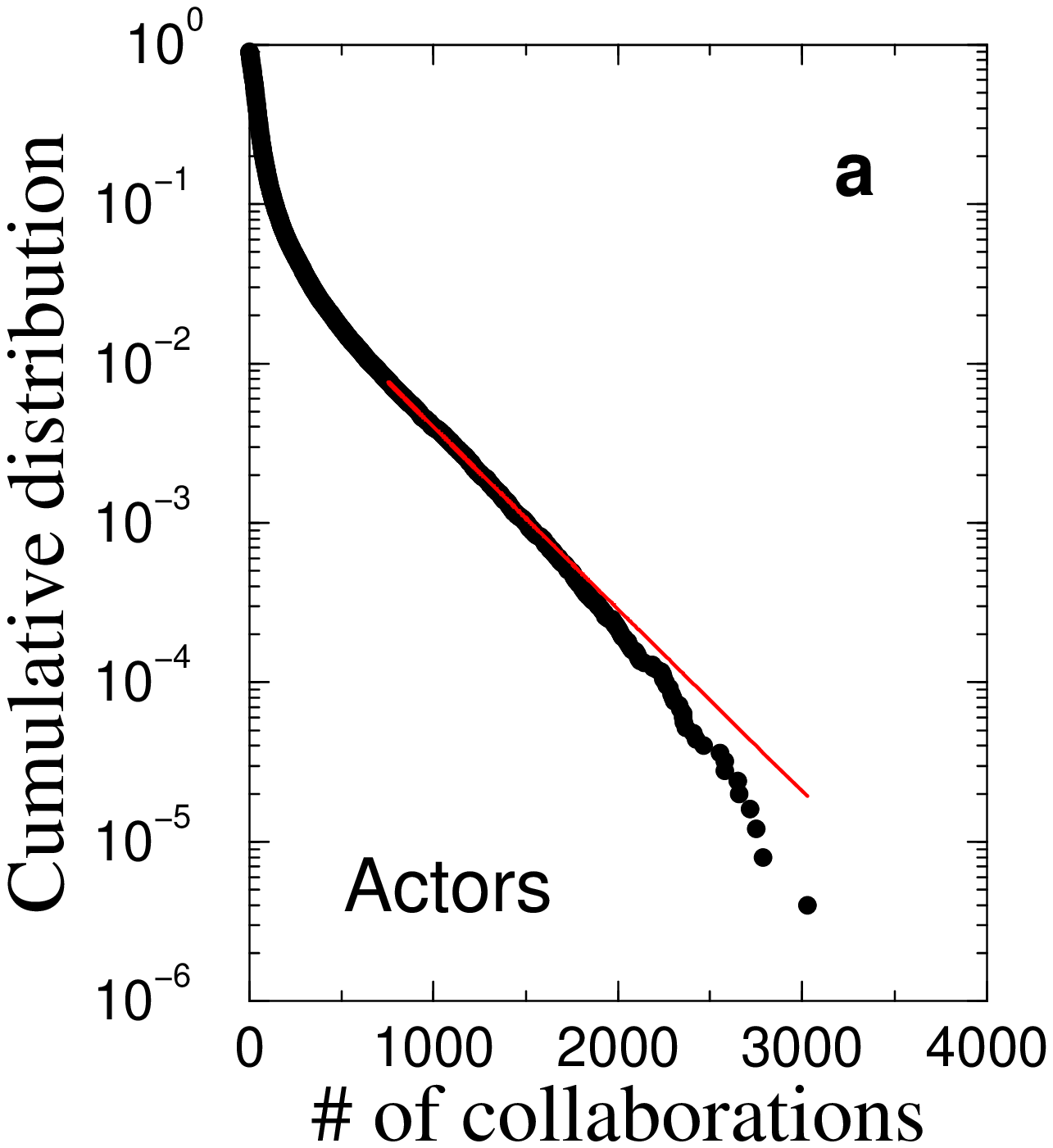}}}
\hspace*{-1.5cm}
\epsfysize=0.4\columnwidth{{\epsfbox{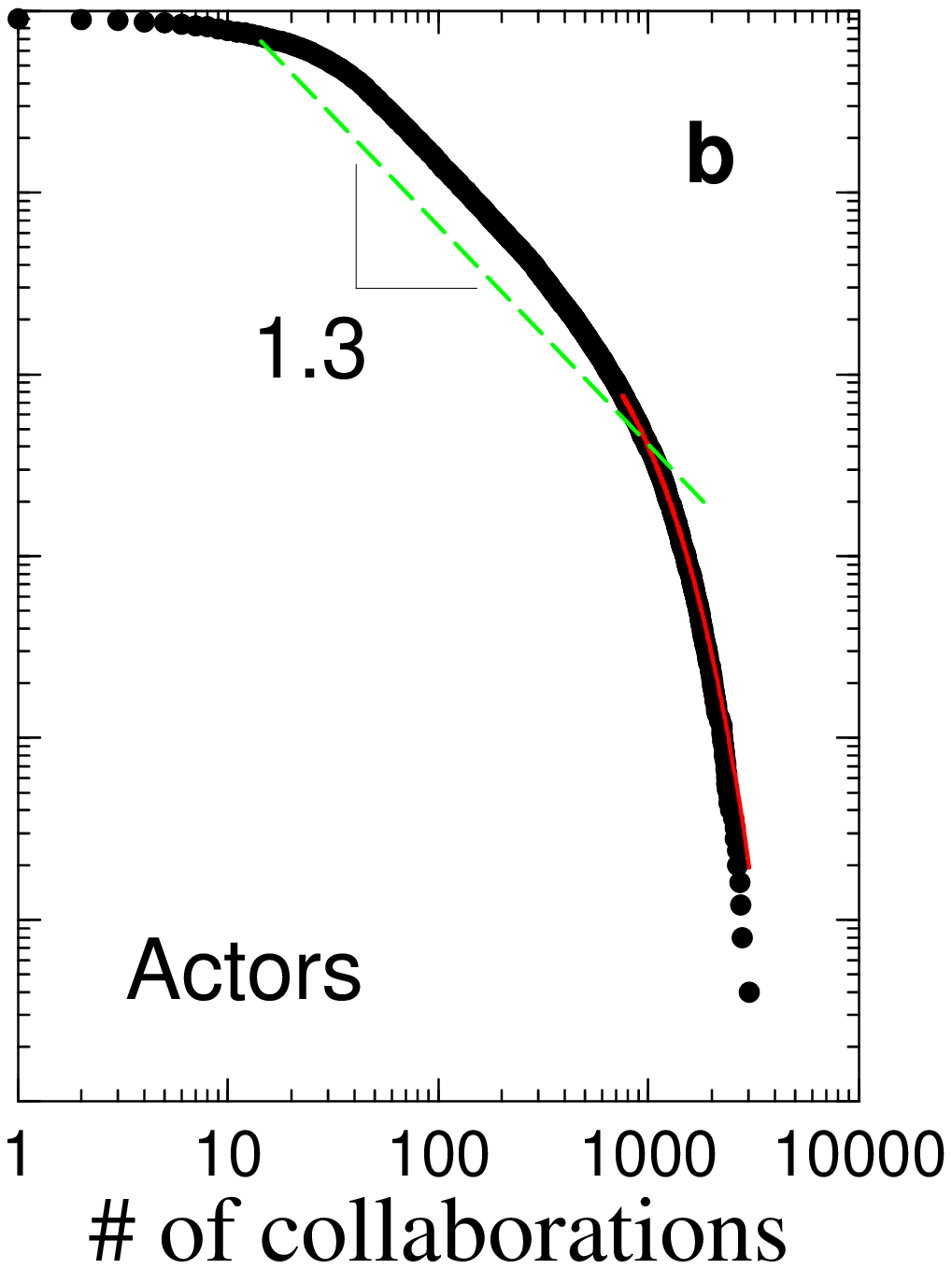}}}
}
\centerline{
\epsfysize=0.4\columnwidth{{\epsfbox{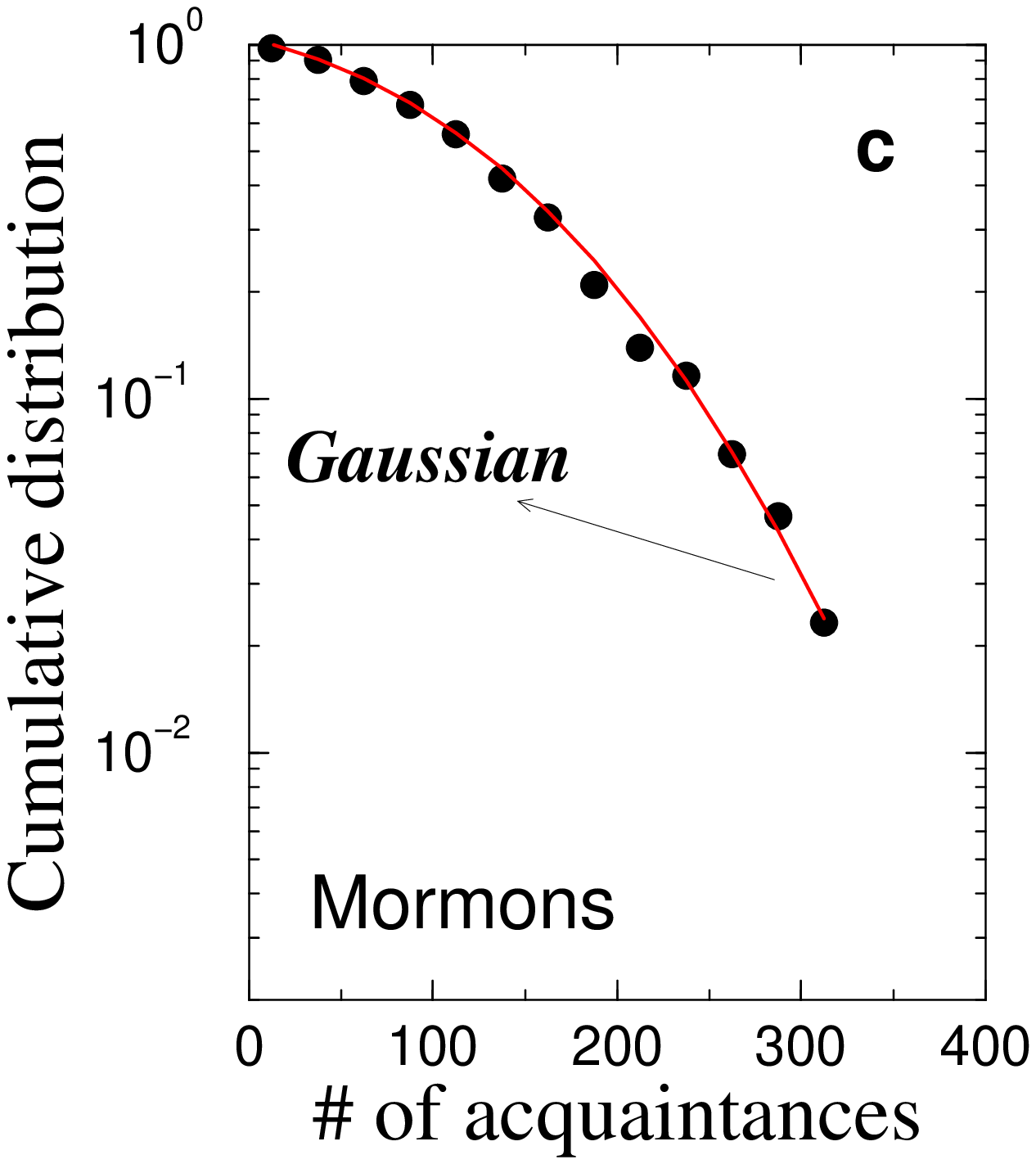}}}
\hspace*{-1.5cm}
\epsfysize=0.4\columnwidth{{\epsfbox{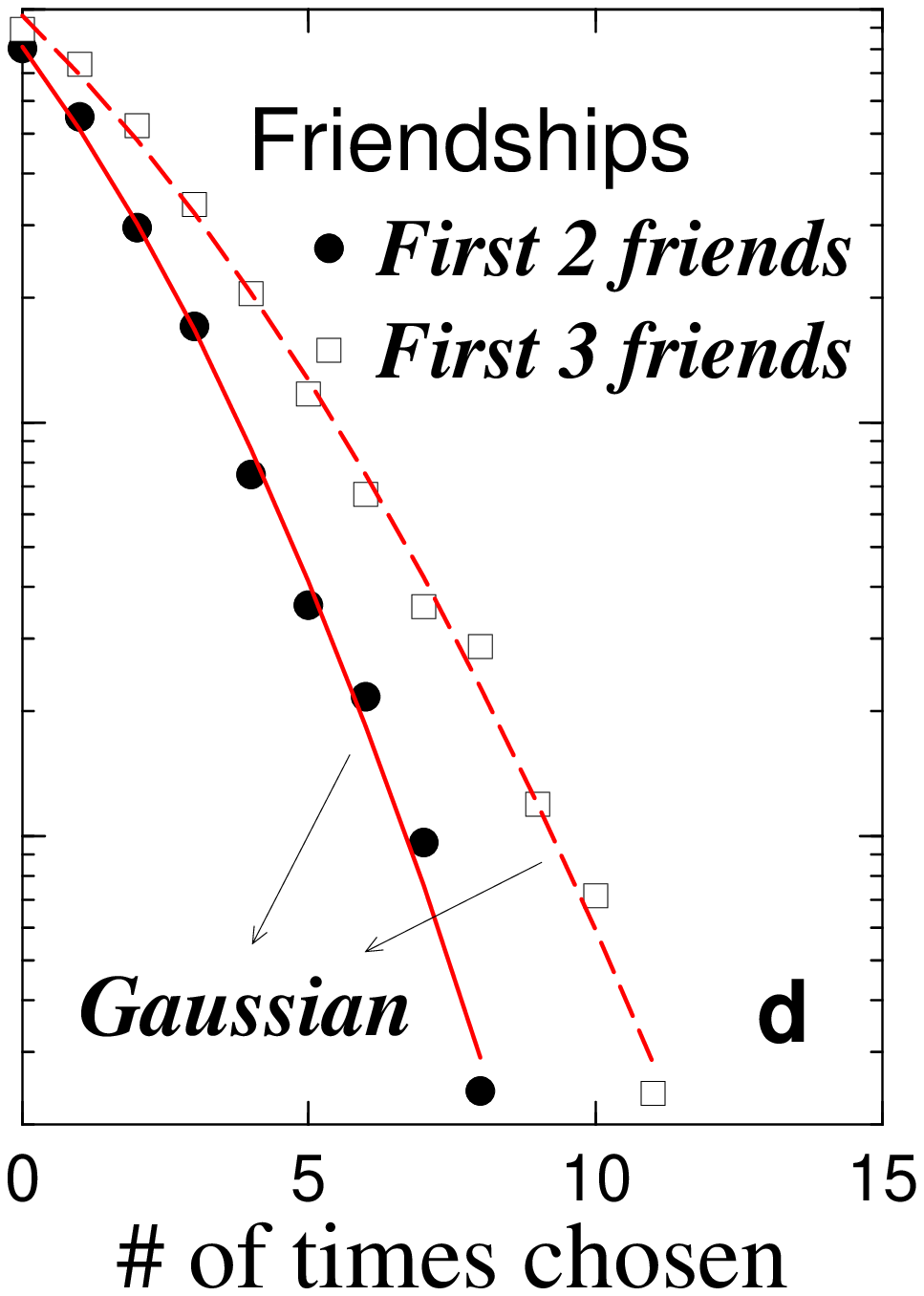}}}
}
\vspace*{.5cm}
\caption{ Social networks. {\bf a} Linear-log plot of the cumulative
distribution of connectivities for the network of movie actors
\protect\cite{Watts98}.  The full line is a guide for the eye of what
an exponential decay would be.  The data appear to fall faster in the
tail than for an exponential decay, suggesting a Gaussian decay.  Both
exponential and Gaussian decays indicate that the connectivity
distribution is not scale free. {\bf b} Log-log plot of the cumulative
distribution of connectivities for the network of movie actors.  This
plot suggests that for values of number of collaborations between 30
and 300 the data are consistent with a power-law decay.  The apparent
exponent of this cumulative distribution, $\alpha - 1 \approx 1.3$, is
consistent with the value $\alpha = 2.3 \pm 0.1$ reported for the
probability density function \protect\cite{Barabasi99b}.  For larger
numbers of collaborations the power-law decays is truncated.  {\bf c}
Linear-log plot of the cumulative distribution of connectivities for
the network of acquaintances of 43 Utah Mormons
\protect\cite{Bernard98}.  The full line is the fit to the cumulative
distribution of a Gaussian.  The tail of the distribution appears to
fall off as a Gaussian, suggesting that there is a single scale for
the number of acquaintances in social networks.  {\bf d} Linear-log
plot of the cumulative distribution of connectivities for the
friendship network of 417 high-school students
\protect\cite{Fararo64}. The number of links is the number of times a
student is chosen by another student as one of his/hers two (three)
best friends. The lines are Gaussian fits to the empirical
distributions.  }
\label{f.social}
\end{figure}

Second, we consider three examples of ``social'' networks: (iii) the
movie-actors network~\cite{Watts98}, the links in this network
indicating that the two actors were cast at least once in the same
movie, (iv) the acquaintance network of Mormons~\cite{Bernard98}, the
vertices being 43 Utah Mormons and the number of links the number of
other Mormons they know, and (v) the friendship-network of 417 Madison
Junior High School students~\cite{Fararo64}. Figure~\ref{f.social}
shows the connectivity distribution for these social networks.  The
scale-free (power-law) behavior of the actor's
network~\cite{Barabasi99b} is truncated by an exponential tail.  In
contrast, the network of acquaintances of the Utah Mormons and the
friendship network of the high-school students display no power-law
regime, but instead we find results consistent with a Gaussian
distribution of connectivities, indicating the existence of a single
scale for $k$.

\begin{figure}
\centerline{\bf FIGURE 3}
\vspace*{1cm}
\centerline{
\epsfysize=0.4\columnwidth{{\epsfbox{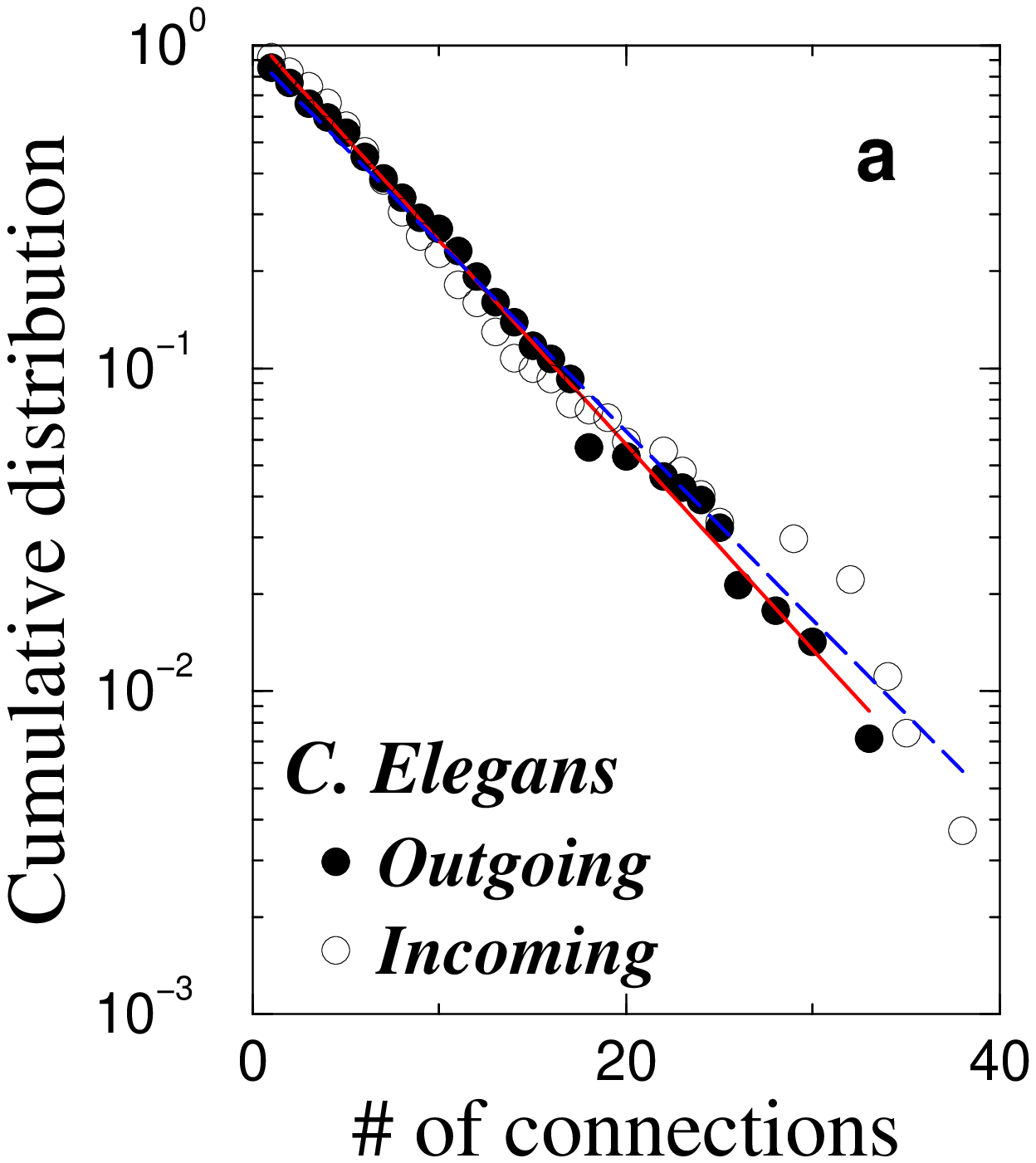}}}
\hspace*{-1.5cm}
\epsfysize=0.4\columnwidth{{\epsfbox{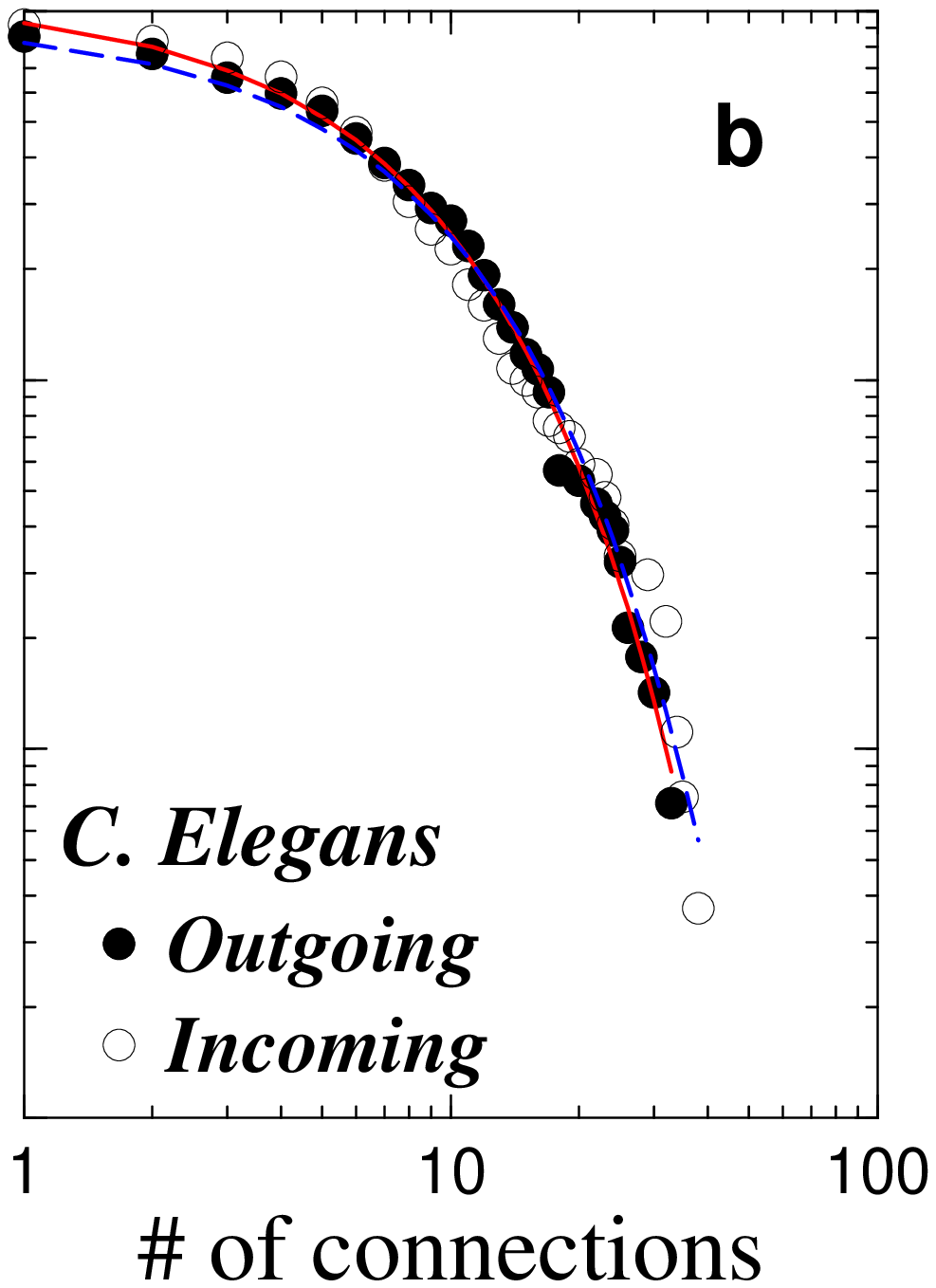}}}
}
\centerline{
\epsfysize=0.4\columnwidth{{\epsfbox{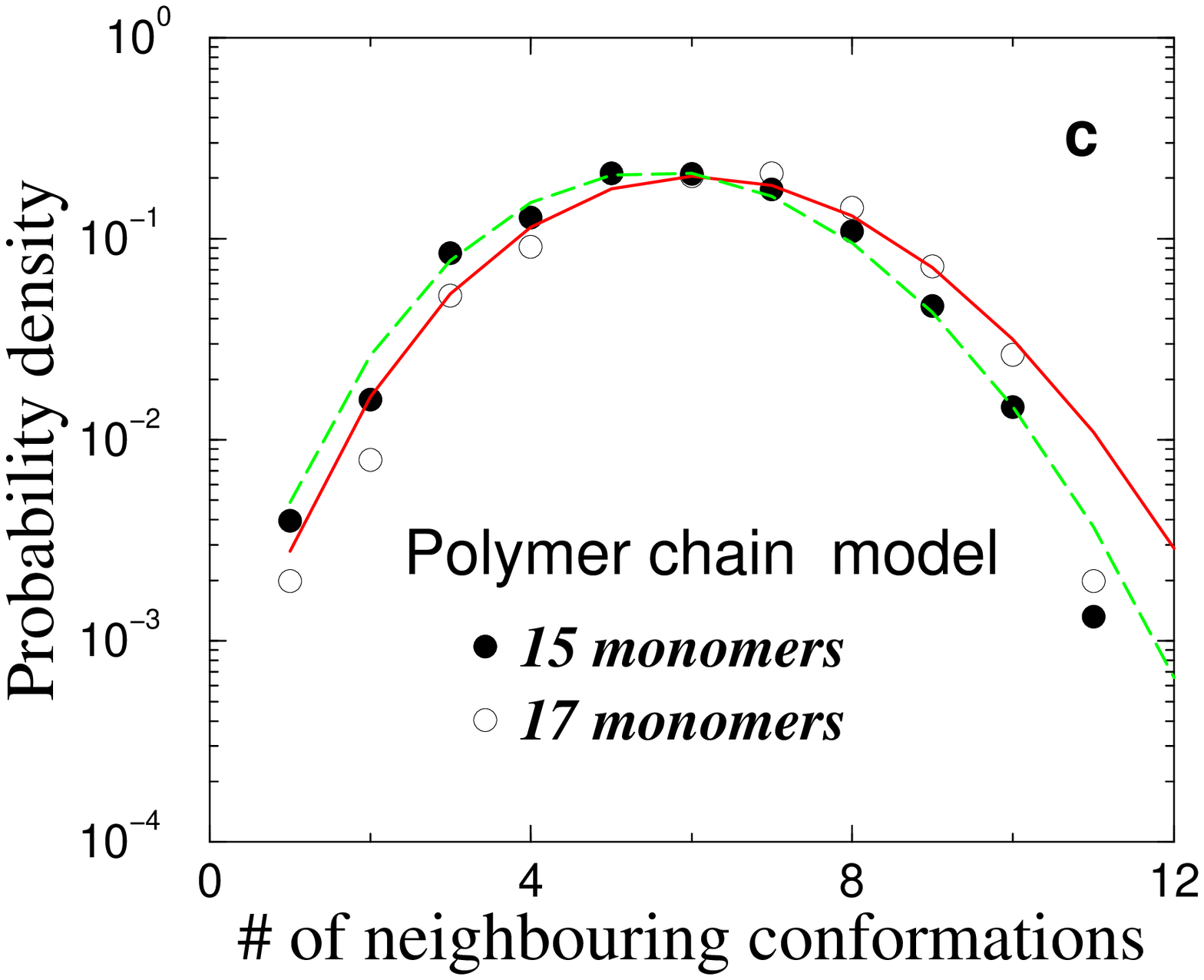}}}
}
\vspace*{.5cm}
\caption{ Biologic and physical networks. {\bf a} Linear-log plot of
the cumulative distribution of outgoing and incoming connections for
the neuronal network of the worm {\it
C.~Elegans}~\protect\cite{White86,Koch99}.  The full and long-dashed
lines are exponential fits to the distributions of outgoing and
incoming connections, respectively.  The tails of the distributions
appear to be consistent with an exponential decay. {\bf b} Log-log
plot of the cumulative distribution of outgoing and incoming
connections for the neuronal network of the worm {\it C.~Elegans}.  If
the distribution would have a power law tail then it would fall on a
straight line in a log-log plot.  The data appear to reject the
hypothesis of a power-law distribution for the connectivity.  {\bf c}
Linear-log plot of the probability density function of connectivities
for the network of conformations of a lattice polymer
chain~\protect\cite{Scala}.  A simple argument suggests that the
connectivities follow a binomial distribution.  The full and dashed
lines are fits of a binomial probability density function to the data
for polymer chains of different lengths.  For the values of the
parameters obtained in the fit, the binomial closely resembles the
Gaussian indicating that there is a single scale for the
connectivities of the conformation space of polymers.  }
\label{f.natural}
\end{figure}

Third, we consider two examples of networks from the natural sciences:
(vi) the neuronal network of the worm {\it C.~Elegans\/}
\cite{Watts98,White86,Koch99}, the vertices being the neurons and the
links being connections between neurons, and (vii) the conformation
space of a lattice polymer chain~\cite{Scala}, the vertices being the
possible conformations of the polymer chain and the links the
possibility of connecting two conformations through local movements of
the chain~\cite{Scala}.  The conformation space of a protein chain
shares many of the properties~\cite{Scala} of the small-world networks
of Ref.~\cite{Watts98}.  Figures~\ref{f.natural}a,b show for {\it
C.~Elegans\/} the cumulative distribution of $k$ for both incoming and
outgoing neuronal links.  The tails of both distributions are well
approximated by exponential decays, consistent with a single scale for
the connectivities.  For the network of conformations of a polymer
chain the connectivity follows a binomial distribution, which
converges to the Gaussian~\cite{Scala}, so we also find a single scale
for the connectivity of the vertices~(Fig.~\ref{f.natural}c).


Thus, there is empirical evidence for the occurrence of three classes
of small-world networks: (a) {\it scale-free\/} networks,
characterized by a connectivity distribution with a tail that decays
as a power law~\cite{Barabasi99a,Seglen92,Redner98}; (b) {\it
broad-scale\/} or truncated scale-free networks, characterized by a
connectivity distribution that has a power-law regime followed by a
sharp cut-off, like an exponential or Gaussian decay of the tail [see
example (iii)]; (c) {\it single-scale\/} networks, characterized by a
connectivity distribution with a fast decaying tail, such as
exponential or Gaussian [see examples (i),(ii),(iv-vii)].

A natural question is ``What are the reasons for such a rich range of
possible structures for small-world networks?''  To answer this
question let us recall that preferential attachment in growing
networks gives rise to a power-law distribution of
connectivities~\cite{Barabasi99b}.  However, preferential attachment
can be hindered by two classes of factors: (I) {\it aging\/} of the
vertices.  This effect can be pictured for the network of actors: in
time, every actress or actor will stop acting.  For the network, this
fact implies that even a very highly connected vertex will,
eventually, stop receiving new links.  The vertex is still part of the
network and contributing to network statistics, but it no longer
receives links. The aging of the vertices thus limits the preferential
attachment preventing a scale-free distribution of connectivities.
(II) {\it cost\/} of adding links to the vertices or the limited {\it
capacity\/} of a vertex.  This effect is exemplified by the network of
world airports: for reasons of efficiency, commercial airlines prefer
to have a small number of hubs where all routes would connect.  To
first approximation, this is indeed what happens for individual
airlines, but when we consider all airlines together, it becomes
physically impossible for an airport to become a hub to all airlines.
Due to space and time constraints, each airport will limit the number
of landings/departures per hour, and the number of passengers in
transit.  Hence, physical costs of adding links and limited capacity
of a vertex \cite{Bonney56,Moreno56} will limit the number of possible
links attaching to a given vertex.

\begin{figure}
\centerline{\bf FIGURE 4}
\vspace*{1cm}
\centerline{
\epsfysize=0.4\columnwidth{{\epsfbox{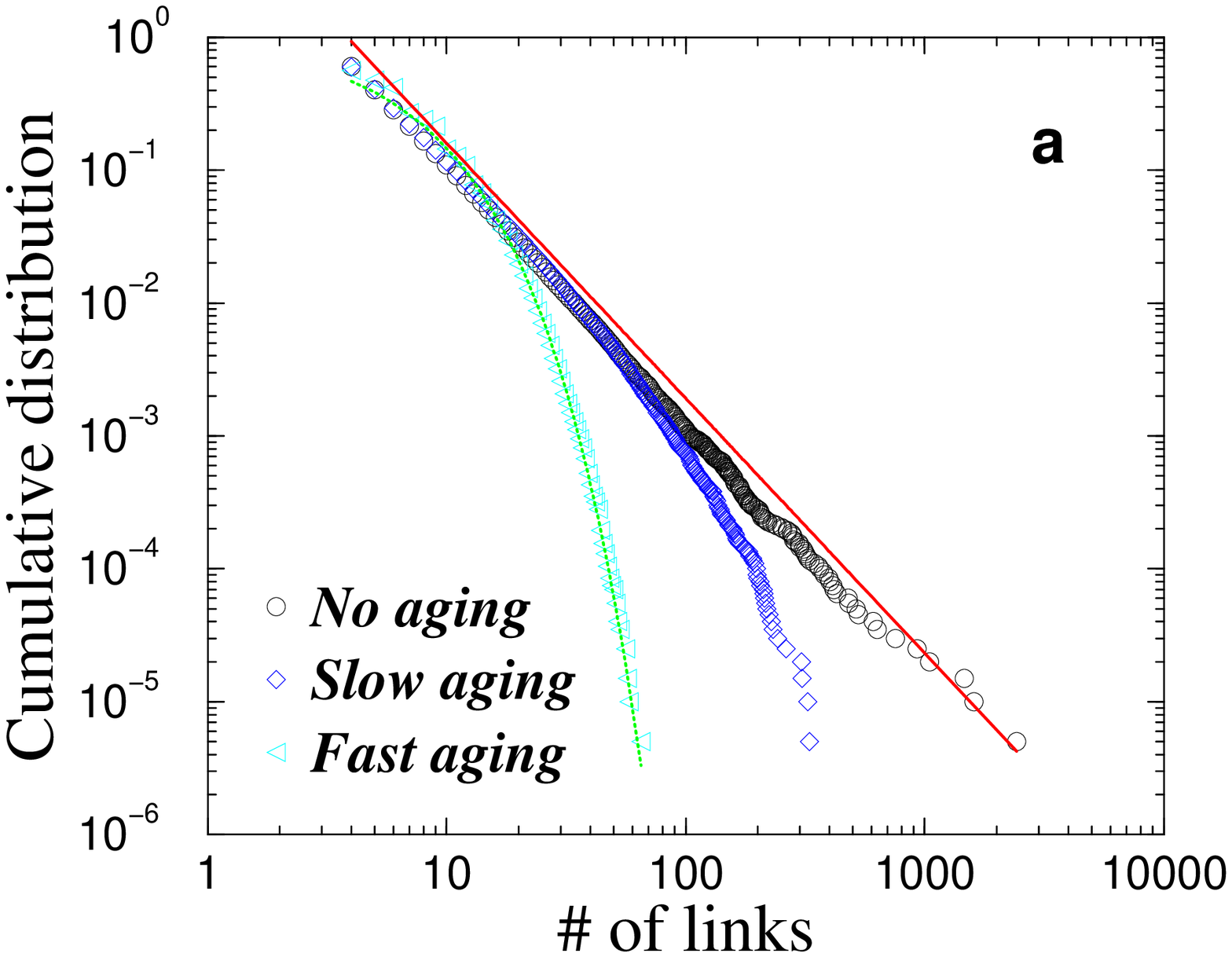}}}
}
\centerline{
\epsfysize=0.4\columnwidth{{\epsfbox{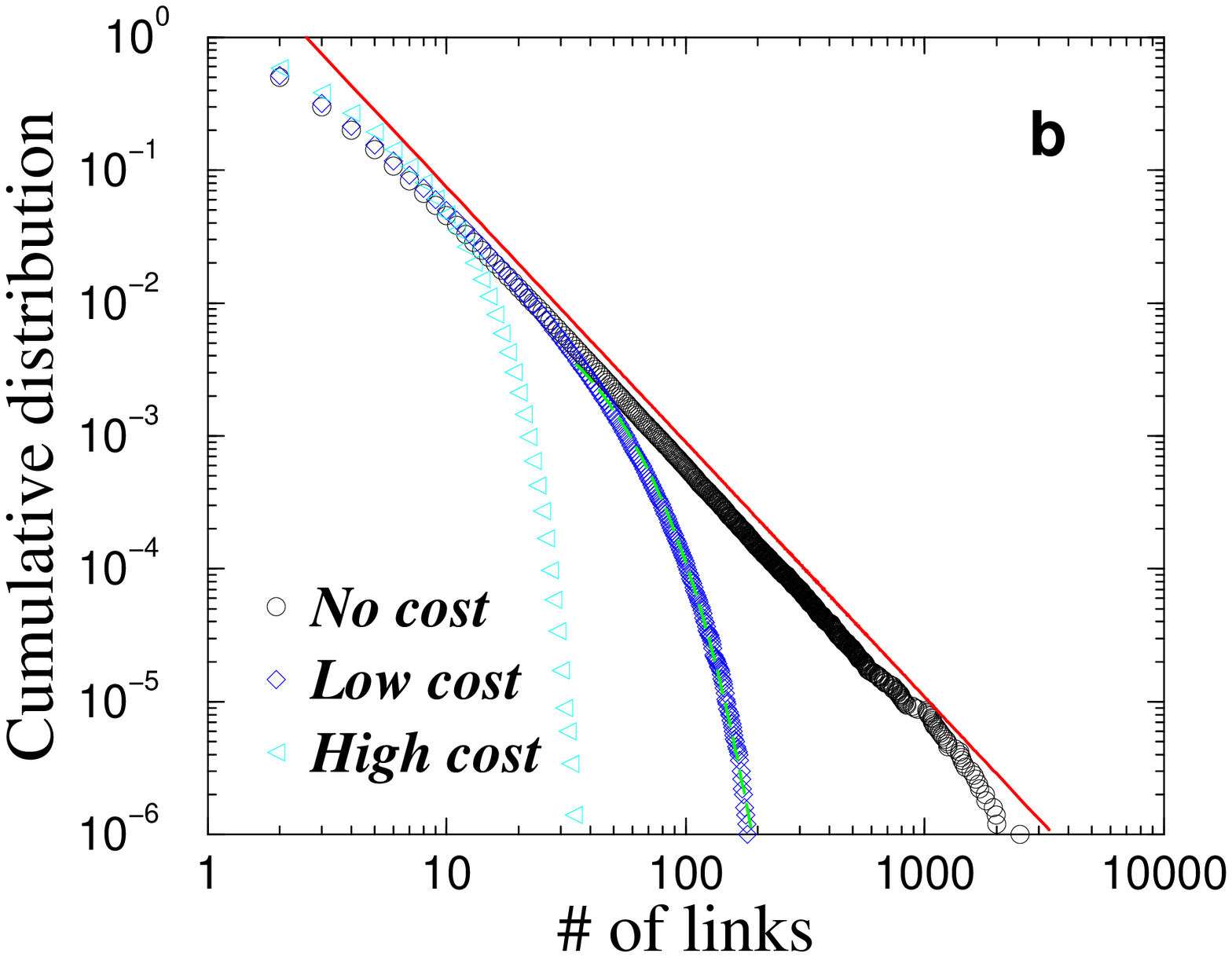}}}
}
\vspace*{.5cm}
\caption{ Truncation of scale-free connectivity by adding constraints
to the model of Ref.~\protect\cite{Barabasi99b}. {\bf a} Effect of
aging of vertices on the connectivity distribution; we see that aging
leads to a cut-off of the power-law regime in the connectivity
distribution.  For sufficient aging of the vertices, the power-law
regime disappears altogether.  {\bf b} Effect of cost of adding links
on the connectivity distribution.  Our results indicate that costs for
adding links also leads to a cut-off of the power-law regime in the
connectivity distribution and that for sufficient costs the power-law
regime disappears altogether.  }
\label{f.cutcut}
\end{figure}

To test numerically the effect of aging and cost constraints on the
local structure of networks with preferential attachment, we simulate
the scale-free model of Ref.~\cite{Barabasi99b} but introduce aging
and cost constraints of varying strength. Figure~\ref{f.cutcut} shows
that both types of constraints lead to cut-offs on the power-law decay
of the tail of connectivity distribution and that for strong enough
constraints no power-law region is visible.


We note that the possible distributions of connectivity of the
small-world networks have an analogy in the theory of critical
phenomena~\cite{Stanley71}.  At the gas-liquid critical point, the
distribution of sizes of the droplets of the gas (or of the liquid) is
{\it scale-free\/}, as there is no free-energy cost in their
formation~\cite{Stanley71}.  As for the case of a scale-free network,
the size $s$ of a droplet is power-law distributed: $p(s) \sim
s^{-\alpha}$.  As we move away from the critical point, the appearance
of a non-negligible surface tension introduces a free-energy cost for
droplets which limits their sizes, so that their distribution becomes
{\it broad-scale\/}: $ p(s) \sim s^{-\alpha} f (s / \xi ) $, where
$\xi$ is the typical size for which surface tension starts to be
significant and the function $f(s / \xi)$ introduces a sharp cut-off
for droplet sizes $s > \xi$.  Far from the critical point, the scale
$\xi$ becomes so small that no power-law regime is observed and the
droplets become {\it single-scale\/} distributed: $p(s) \sim f (s /
\xi)$.  Often, the distribution of sizes in this regime is exponential
or Gaussian.

We thank J.S.~Andrade~Jr., R.~Cuerno, N.~Dokholyan, P.~Gopikrishnan,
C.~Hartley, E.~LaNave, K.B.~Lauritsen, H.~Orland, F.~Starr and
S.~Zapperi for stimulating discussions and helpful suggestions.  The
Center for Polymer Studies is funded by NSF.



\end{multicols}


\newpage

\newpage

\newpage

\newpage

\end{document}